\def\year{2022}\relax
\documentclass[letterpaper]{article} 
\usepackage{aaai22}  
\usepackage{times}  
\usepackage{helvet}  
\usepackage{courier}  
\usepackage[hyphens]{url}  
\usepackage{graphicx} 
\usepackage{natbib}  
\usepackage{caption} 
\usepackage{xcolor}
\DeclareCaptionStyle{ruled}{labelfont=normalfont,labelsep=colon,strut=off} 
\usepackage{algorithm}
\usepackage{algorithmic}

\usepackage{newfloat}
\usepackage{listings}
\floatstyle{ruled}
\newfloat{listing}{tb}{lst}{}
\floatname{listing}{Listing}
\usepackage{dirtytalk}
\usepackage{booktabs}
\usepackage{tabularx}
\usepackage{multirow}
\usepackage{color,soul}
\usepackage{subcaption}

\usepackage{hyperref}

\title{Online search is more likely to lead youth to validate true news than to refute false ones}
\author{
    Azza Bouleimen\textsuperscript{\rm 1, 2},
    Luca Luceri\textsuperscript{\rm 3},
    Felipe Cardoso\textsuperscript{\rm 1},
    Luca Botturi\textsuperscript{\rm 1},
    Martin Hermida\textsuperscript{\rm 4},
    Loredana Addimando\textsuperscript{\rm 1},
    Chiara Beretta\textsuperscript{\rm 1},
    Marzia Galloni\textsuperscript{\rm 1},
    Silvia Giordano\textsuperscript{\rm 1}
}
\affiliations{
    \textsuperscript{\rm 1}University of Applied Sciences and Arts of Southern Switzerland - SUPSI\\
    \textsuperscript{\rm 2} University of Zurich - UZH\\
    \textsuperscript{\rm 3} Information Sciences Institute - University of Southern California\\
    \textsuperscript{\rm 4} The Schwyz University of Teacher Education - PHSZ\\
    azza.bouleimen@supsi.ch

}

\begin{document}

\maketitle

\begin{abstract}
With the spread of high-speed Internet and portable smart devices, the way people access and consume information has drastically changed. However, this presents many challenges, including information overload, personal data leakage, and misinformation diffusion. Across the spectrum of risks that Internet users face nowadays, this work focuses on understanding how young people perceive and deal with false information. Within an experimental campaign involving 183 students, we presented six different news items to the participants and invited them to browse the Internet to assess the veracity of the presented information. Our results suggest that online search is more likely to lead students to validate true news than to refute false ones. We found that students change their opinion about a specific piece of information more often than their global idea about a broader topic. Also, our experiment reflected that most participants rely on online sources to obtain information and access the news, and those getting information from books and Internet browsing are the most accurate in assessing the veracity of a news item. This work provides a principled understanding of how young people perceive and distinguish true and false pieces of information, identifying strengths and weaknesses amidst young subjects and contributing to building tailored digital information literacy strategies for youth.
\end{abstract}

\vspace{-20pt}
\section{Introduction}
During the past two decades, our society has been experiencing rapid changes and multiple challenges tied to significant innovations in information technology. In particular, access to the Internet has skyrocketed and diffused on a global scale: As of April 2022, 5 billion people are using the Internet, and 93\% of them are social media users~\footnote{\href{https://www.domo.com/data-never-sleeps}{https://www.domo.com/data-never-sleeps}. Accessed November 6\textsuperscript{th}, 2023.}, leading to exposure to a tremendous amount of information on a daily basis. This has brought many positive impacts to societies, for instance, a rise in awareness about social and political issues, which also generated increased activism among citizens ~\cite{enikolopov2020social}. However, in recent years, this abundance in information and broadcasting tools has led to a concerning phenomenon referred to as \textit{misinformation}, which is defined as the spreading of false or misleading information~\cite{lazer2018science}.

Fake or manipulated content is nothing new: for instance, made-up press stories about purported life on the moon were already published in 1835~\cite{pennycook2021psychology}. Nevertheless, the Internet, and particularly social media, gave this phenomenon an unprecedented size and impact. Concerns about misinformation have become particularly important when orchestrated fake news campaigns were performed on social media to manipulate public opinion during the Brexit Referendum, US Elections, and other political events~\cite{budak2019happened,guess2018selective,bessi2016social,suresh2023tracking,luceri2023unmasking}. Alarmingly, the spread of misinformation and hateful content on Facebook led to major killings, human rights violations, and war crimes in Myanmar between October 2016 and January 2017~\cite{yue2019weaponization}. Misinformation has also proliferated in the public health sphere during the Covid-19 outbreak ~\cite{ferrara2020misinformation, nogara2022disinformation} when a fake piece of information caused the death of 800 people, drove 5876 people to hospitals, and caused full blindness to 60 other individuals~\cite{islam2020covid}. Information and misinformation also play a crucial role in geopolitical events~\cite{pierri2022does,ezzeddine2023exposing} as during the ongoing wars in Ukraine~\cite{pierri2023propaganda} and Palestine~\cite{dey2024coordinated}.

Addressing misinformation from a research point of view could take different perspectives. In this paper, we look at the misinformation phenomenon from the angle of Information Literacy (IL). IL is the umbrella concept that collects the competencies required to effectively and efficiently retrieving information, critically selecting and evaluating the sources, and behaving ethically when sharing information~\cite{unesco2003prague}. In this regard, we are interested in understanding factors impacting the truth assessment capabilities of users online, in particular young people. 
In fact, responsibly acknowledging the threat of misinformation and its rapidly changing manifestations means heading the call for educating the young generations and providing them with the means for correctly assessing information online. Especially considering the vulnerable situation their young age and lack of experience are likely to expose them to. In addition, research suggests an important change in how young people use the Internet to explore new content or find information. For instance, a Google internal research~\cite{perez2022google}, involving American participants aged 18-24, found that younger users often turn to social media apps such as Instagram or TikTok to discover new content using uncommon queries. These changes in the behavior among young Internet users are an additional motivation to tailor our study specifically around young people. Hence, we explore how they cope with misinformation and search for accurate information online.

\paragraph{Our contributions.}
In this work, we intend to study young people's behavior when confronted with a piece of supposedly true or false information on the Internet. To this aim, we perform an experiment asking the participants to evaluate the accuracy of several \textit{news items}. Specifically, during our experiment, we present six news items to the participants, and we investigate their perception of the accuracy of the news items - namely \textit{accuracy judgment}, as referred to in~\cite{pennycook2021psychology}, before and after an online search. We also examine the impact of the search on their opinion about the general topic related to the news (from now on, \textit{general opinion}). 
To have a principled understanding of how youth deal with questionable information and examine the factors potentially affecting their accuracy judgment, we aim to answer the following research questions (RQs): 
\begin{itemize}
    \item[] \textbf{RQ1:} Does the accuracy judgment of a news item change before and after carrying out an online search? Does this change correlate with participants’ change in their general opinion?
    \item[] \textbf{RQ2:} How accurate are the participants in discerning true and false news items? Can we identify characteristics of the participants tied to their capacity to classify news items?
    \item[] \textbf{RQ3:} Is there any relation between the information source used to verify the information and the correct assessment of a news item? Do participants change their preferred information source after the online search?
\end{itemize}

To answer these questions, we designed an experiment involving 261 students from high school and university. During the experiment, we asked them questions about specific topics, presented them with six news items, and invited them to browse the Internet to assess the information's veracity. The data collected from this experiment are available to the research community\footnote{\url{https://osf.io/mejfn/}}.

\vspace{-10pt}
\section{Related work}
\label{RW}
Our study falls at the intersection of research areas such as Information Literacy, Computer Science, and Psychology. We describe the related literature in the following sections.
\vspace{-8pt}
\subsection{Information Literacy}

Dealing with fake news and false information is a key aspect of IL. All current Digital and Media Literacy models such as DigComp 2.1~\cite{carretero2017digital} include IL, which is also regarded as a central element in school education~\cite{bucher2000importance}; and indeed, today, IL is included in many compulsory education school curricula. IL can also be considered essential to establish an effective individual lifelong learning strategy~\cite{kurbanoglu2013analysis}. In fact, evidence suggests that IL competencies enable users to better recognize fake news~\cite{Jones_Jang_2021_media}, and correlate higher with information search capacities than does digital nativity~\cite{ccoklar2017information}. This highlights the paramount importance of educating young people on IL.
\vspace{-8pt}
\subsection{Computer Science \& Psychology}
There exists established literature on studying the spread of fake news on social media~\cite{vosoughi2018spread, Fake_news_Twitter_2016_US_election} and the characterization of user profiles that are vulnerable to fake news~\cite{ye2023susceptibility,guess2018selective, pennycook2020falls}. For instance, the work of~\cite{vosoughi2018spread} studies the spread of true and false news online from $\sim$126,000 claims of English-text tweets between 2006 and 2017. The claims were shared by $\sim$3 million people more than 4.5 million times. The authors discovered that false news spread significantly farther, faster, deeper, and more broadly than truth. They support the idea that true news takes six times as long as false information to reach 1,500 people and twenty times longer to be shared ten times from the original tweet. They argue that falsehoods are 70\% more likely to be retweeted than the truth. More recent studies examine the impact of prior exposures on users' susceptibility to misinformation~\cite{ye2023susceptibility} or investigate cognitive and psychological aspects of users susceptible to fake news~\cite{pennycook2020falls}. \cite{ye2023susceptibility} provided evidence indicating that greater exposure significantly enhances the probability of adoption, revealing that users are more inclined to adopt sources of low credibility after fewer exposures compared to sources of high credibility. \cite{pennycook2020falls} found that the tendency to ascribe profundity to randomly generated sentences correlates positively with the perception of fake news and that people who overclaim their level of knowledge also judge fake news to be more accurate. Other research ~\cite{Fake_news_Twitter_2016_US_election, chen2020neutral,luceri2019red,luceri2021down} found that political leaning plays a relevant role in misinformation adoption, with conservative individuals more likely to engage with and share low-credibility information on social media. 

Another line of research focuses on actionable strategies to reduce belief in false narratives~\cite{morrow2022emerging,yaqub2020effects} and interventions to curb misinformation spread on social media~\cite{lu2022effects}. For example, in~\cite{pennycook2020fighting,pennycook2021shifting}, the authors suggest that nudging people to consider accuracy on social media platforms reduces sharing misinformation online. In the review conducted in~\cite{martel2023misinformation}, the authors find that misinformation warning labels are widely effective in preventing individuals from adopting false narratives. In a related experiment on the relation between the novelty of news and the effectiveness of labeling articles as more or less accurate~\cite{nevo2022topic}, the authors find that novelty influences the effectiveness of labeling articles: news readers are more likely to change their judgment if they are less familiar with the topic.
\cite{lewandowsky2012misinformation} studied the complex psychological mechanism involved in hindering misinformation correction. They present some recommendations for debunking myths, such as repeated retractions, avoiding making people more familiar with misinformation and using simple and brief argumentation to refute the false information. 

Our research takes inspiration from~\cite{pennycook2020fighting,pennycook2021shifting}, where researchers found that participants were less able to discern a news item (true vs. false) when they were asked about whether they would share a headline on social media rather than directly asking them about its accuracy. The present study tries to add new insights to the existing corpus from at least two points of view. First, while the studies mentioned above mainly involved North American adults, ours carried out an experiment specifically designed for young people from Italy and Switzerland, from which we obtained an original dataset focusing exclusively on youth. Second, our experiment stands out from other fake news-related studies because it sheds light on the impact of online search on the accuracy judgment of a multifaceted set of news items by also considering the role of several online and offline information sources. 

\vspace{-10pt}
\section{Experiment \& Data}
\label{data}
This work is based on an experiment composed of two parts, namely the \textit{profile survey} and the \textit{task survey}. 

In the first part, we provide a \textit{profile survey} to the participants, which aims at (a) collecting their demographic data (e.g., age, gender, school grade, socio-economic level) and (b) conducting a rational thinking assessment leveraging the Rational Thinking Level \cite{norris2011experiential} and Cognitive Reflection Test \cite{frederick2005cognitive} scales. In the second part, we present a \textit{task survey} to the participants in which they are asked to answer some questions and perform an online search. The task survey is composed of six different independent tasks. Every task refers to a topic, as presented in Table~\ref{tab:event}. The choice of the topics was made in a way to alternate between \say{Serious} and \say{Leisure} topics to potentially meet different participant's interests. For every topic, we propose a (false or true) news item presented in the form of a headline composed of a title, caption, and picture, similarly to~\cite{pennycook2020fighting}. The interested reader can refer to the \textit{Appendix} for a detailed overview of the news items. Every task can be performed only once, and every time the participants log in, the six tasks are presented in a random order. 
Participants performed the whole experiment (including the six tasks) from their own devices and using a dedicated Web platform, which integrated surveys implemented with Qualtrics. It is important to note that the participants were told that they took part in an experiment about their online search behavior and were not aware of the scope of the experiment in relation to fake news. 

\begin{table}[h]
    \centering
  \caption{Summary of the six different tasks presented to the participants.}
  \label{tab:event}
  \begin{tabular}{p{0.01\columnwidth}p{0.23\columnwidth}p{0.12\columnwidth}p{0.34\columnwidth}p{0.1\columnwidth}}
    \toprule
    \textbf{n°}&\textbf{Topic}&\textbf{Type}&\textbf{News item}&\textbf{Veracity}\\
    \midrule
1  & Viruses             & Serious       & Coconut oil against Covid-19 & False             \\
2  & Human Rights        & Serious       & Detention camps in Canada    & False             \\
3  & Asian Food          & Leisure       & McDog in Korea               & False             \\
4  & Climate Change      & Serious       & Climate change and Malaria   & True              \\
5  & Rights of Prisoners & Serious       & Baby Shark torture           & True              \\
6  & Cheating in exams   & Leisure       & Driver license fraud         & True  \\           
  \bottomrule
\end{tabular}
\end{table}

Every task consists of three phases, namely \say{Background}, \say{Reaction}, and \say{Perception}. 
During the \textit{Background} phase, we examine to what extent the participants are acquainted with the topic of a given task without specifically examining their opinion on the corresponding news item. Specifically, we ask for information about their prior knowledge of and interest in the topic, as well as which sources they rely on when they look for information about the topic. These information sources can be: \say{Internet browsing}, \say{newspapers (online and on paper)}, \say{social media}, \say{books}, \say{radio}, \say{TV}, \say{other} or \say{I didn't get/look for this information}. Even though the experiment is about online search, we aim to have a comprehensive overview of the media diet of youth. This will enable us to locate better online search practices of participants relative to all possible information sources~\cite{moody2009constructivist}. After completing the \textit{Background} phase, we present the news item to the participants. Taking inspiration from~\cite{pennycook2020fighting}, the news item is displayed on the dedicated platform in the form of a social media post, with a headline, a photo, and a lead (An example is provided in the \textit{Appendix}). Note that participants cannot click on the presented post, and related headline, and no full article is made available.


In the \textit{Reaction} phase, we investigate participants’ points of view about the presented news item, and we invite them to search online to verify its accuracy. Specifically, we aim to understand whether they would read the corresponding article if they find it on a social media platform, if they would be willing to share the headline on their favorite social network, and how accurate they think the presented headline is. At the end of the Reaction phase, we ask: \say{If you have not done it yet, search this information on the Internet and verify its accuracy.}

Once they complete their online search, they move to the last phase of the survey, namely the \textit{Perception} phase. During this phase, we ask the same questions as the \textit{Background} phase. The purpose is to study whether performing an online search had an impact on the beliefs and behaviors of the participants. Finally, we question them again about the accuracy of the presented headline (same scales and questions as during the reaction phase). A summary of the questions of the task survey is presented in Table~\ref{tab:questions} (Questions on sharing and accuracy are inspired by~\cite{pennycook2020fighting}).

\begin{table*}[h]
  \caption{Summary of the questions of the \textit{task survey}.}
  \label{tab:questions}
  \vspace{-10pt}
\begin{tabular}{p{0.08\textwidth}p{0.1\textwidth}p{0.44\textwidth}p{0.22\textwidth}} 
\toprule
\textbf{Phase}                      & \textbf{Question title}                                                                                                          & \textbf{Question}                                                                                                                                                                                                    & \textbf{Scale / possible answers}                                                                                                                                                                           \\
\midrule
\multirow{4}{*}{Background} & \textit{Knowledge}                                                                                            & 1.1) \say{How do you rate your knowledge about the topic of the task?}.                                                                                                                                & 1 (lowest knowledge) to 5 (highest knowledge)                                                                                                                                             \\
\cmidrule(l){2-4}
                            & \textit{Interes}t                                                                                              & 1.2) \say{How do you rate your interest in this topic?}                                                                                                                                                & 1 (lowest interest) to 5 (highest interest)                                                                                                                                               \\
                            \cmidrule(l){2-4}
                            & \textit{Information Source}                                                                                    & 1.3) \say{Where do you most often get information about this topic?}                                                                                                                                   & \begin{tabular}{@{}p{0.22\textwidth}@{}}Internet browsing, newspapers (online and on paper), social media, books, radio, TV, other, I didn’t get/look for this information.\end{tabular} \\
                            \cmidrule(l){2-4}
                            & \begin{tabular}{@{}p{0.1\textwidth}@{}}\textit{General Opinion}\end{tabular}                & \begin{tabular}{@{}p{0.44\textwidth}@{}}1.4) \say{Do you think that natural remedies can be an effective therapy against viruses?} This is an example of \textit{general opinion} on the topic (virus) related to the news of Task~1 (Covid-19).\end{tabular} & 1 (Definitely Unlikely) to 4 (Definitely Likely).                                                                                                                                       \\
                            \midrule
\multirow{3}{*}{Reaction}  & \textit{Reading}                                                                                               & 2.1) \say{If you were to see this post on social media, would you read the corresponding article?}                                                                                                     & \begin{tabular}{@{}p{0.22\textwidth}@{}}\say{yes}, \say{may be}, \say{no}.\end{tabular}                                                                                                                        \\
\cmidrule(l){2-4}
                            & \textit{Sharing}                                                                                               & 2.2) \say{Would you consider sharing this story on your favorite social network?}                                                                                                                      & \begin{tabular}[c]{@{}p{0.22\textwidth}@{}}\say{yes}, \say{may be}, \say{no}.\end{tabular}                                                                                                                        \\
                            \cmidrule(l){2-4}
                            & \begin{tabular}{@{}p{0.1\textwidth}@{}}\textit{Accuracy Judgment}\end{tabular} & 2.3) \say{To the best of your knowledge, is the claim in the above headline accurate?}.                                                                                                               & 1 (definitely unlikely) to 4 (definitely likely).                                                                                                                                         \\
                            \midrule
Perception                  & \multicolumn{3}{p{0.9\textwidth}}{Questions from the \textit{Background} phase were asked again. Question 1.3 was slightly rephrased as: \say{Where will you search about this topic in the future?}. To conclude, question 2.3 from the \textit{Reaction} phase was presented again to the participants.}    \\
 \bottomrule
\end{tabular}
\end{table*}

The experiment started in November 2020 and involved 261 participants, who were asked to freely complete the profile survey and the task survey with the six information items \cite{botturi2021search}. However, not all participants did the online search when invited to. On average, 39\% of users that answered the task survey of a specific task performed the search. The precise number of participants performing the search on each task can be found in the \textit{Appendix}. Overall, 183 users did at least one task with a search, and 24 users performed all six tasks with an online search. Considering that in this work, we study the impact of online search on telling true from false information, in the following, we only restrict the analysis to the subset of users that did the search for a specific task (183 users).

Participants were recruited from three educational institutions, a public high school from Varese (Italy), SUPSI (Switzerland), and PHSZ (Switzerland), through keynotes at their facilities, as well as via email. Ten gift cards were used as incentives to foster participation in the experiments. 

Overall, 175 participants performed both the profile and the task survey, while 8 participants completed only the task survey. The majority of participants were female (67.2\%), and 88\% of them were under 20 years old. The interested reader can find in the \textit{Appendix} some of the outcomes of the profile survey. In the experiments, 183 participants performed a total of 581 tasks, leading to 581 (\textit{user}, \textit{task}) different instances of participants’ activity in the task survey.

\vspace{-8pt}
\section{Results}
\label{results}

In this section, we describe the main results of our analysis to address the proposed RQs. First, we study the change in the accuracy judgment of both the news item and the topic. Then, we observe whether and how participants correctly identified the news as true or false. Finally, we focus on the resources used by the participants to look for information throughout the experiment, and we investigate whether there is a relation between the information source and their ability to classify the news items accurately. 

\vspace{-10pt}
\subsection{Online search’s impact on accuracy judgment and general opinion (RQ1)}
\subsubsection{Change in accuracy judgment}
\begin{figure*}[t]
\captionsetup[subfigure]{justification=centering} %
    \centering
      \begin{subfigure}{0.49\textwidth}
        \includegraphics[width=\textwidth]{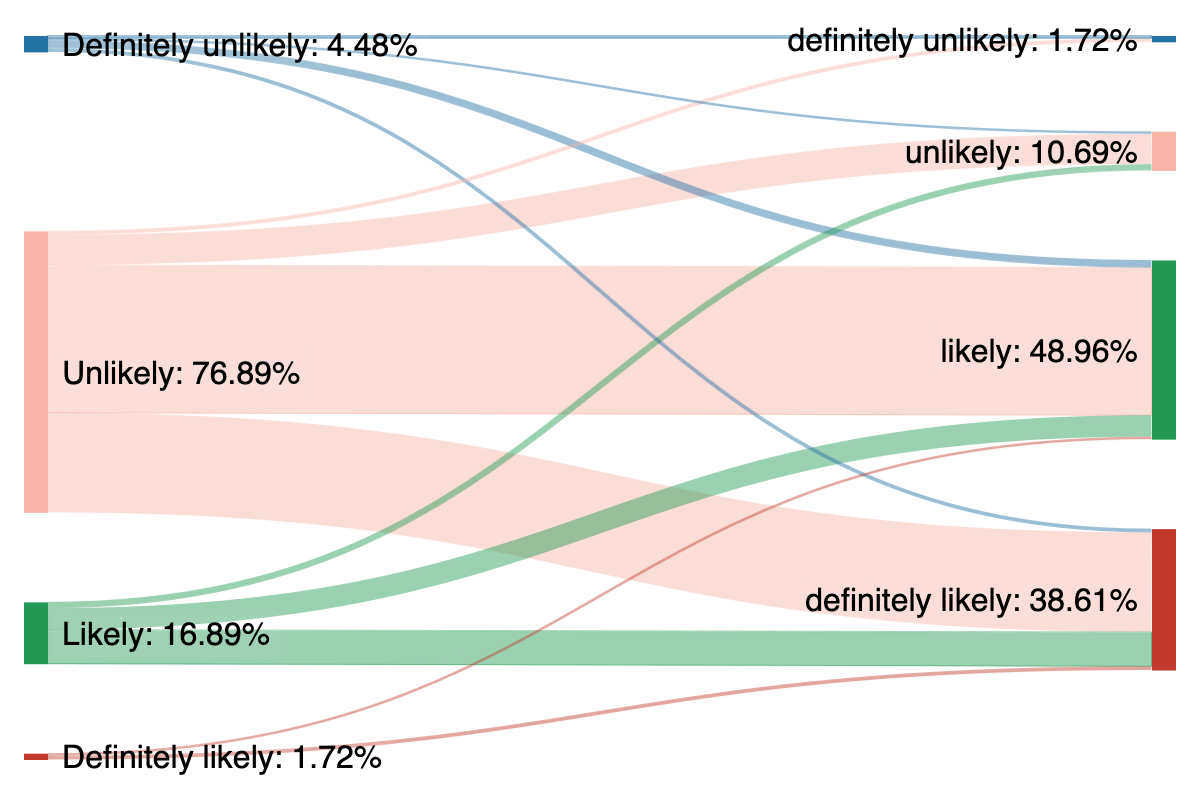}
          \caption{True News}
          \label{accuracy_true}
      \end{subfigure}
      \begin{subfigure}{0.49\textwidth}
        \includegraphics[width=\textwidth]{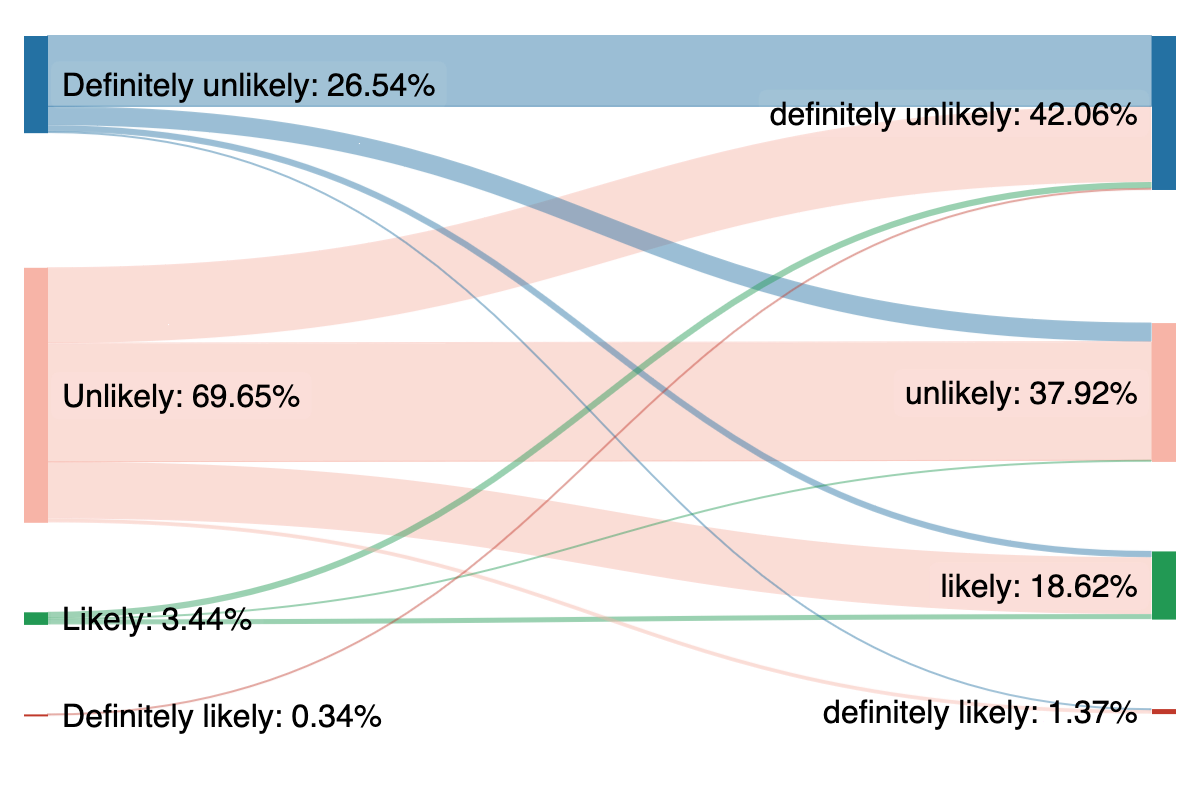}
          \caption{False News}
          \label{accuracy_false}
      \end{subfigure}
      \vspace{-10pt}
\caption[Illustration of the change in accuracy levels]{Illustration of the change in accuracy levels between \textit{Reaction} and \textit{Perception} phases for true and false news. (a) True news: p-value~$<$~0.001. (b) False news: p-value~$=$0.79.}
\label{accuracy_change}
\end{figure*}

To respond to RQ1, we investigate the impact of online search on the accuracy judgment of the presented news items. In Figure~\ref{accuracy_change}, we observe the change of the accuracy levels (ranging from \say{Definitely Unlikely} to \say{Definitely Likely}) for true and false news from the Reaction to the Perception phase. We remind the reader that the same question on accuracy judgment is posed to the participant before and after the online search. On the left (resp. right) side of each chart, we represent the four starting accuracy levels (resp. final levels). The numbers on the chart next to each level indicate the percentage of (\textit{user}, \textit{task}) instances for a given accuracy level. Figure~\ref{accuracy_change} allows us to appreciate the direction of change in the accuracy levels. We denote the direction as \say{correct} if it corresponds to (i) an increase in the accuracy level for true news or (ii) a decrease in the accuracy level for false news. Inversely, the direction is said to be \say{wrong} if it corresponds to (i) a decrease in the accuracy for true news or (ii) an increase in the accuracy level for false news.

From Figure~\ref{accuracy_true}, we observe that before the online search, only about 19\% of the participants correctly considered the presented piece of news as \say{Likely} or \say{Definitely Unlikely}. This proportion increased to about 88\% after the online search. From the chart, we see that many participants who initially assessed the news items as \say{Definitely Unlikely} or \say{Unlikely} (i.e., they thought the news item was false) changed their accuracy judgment after the online search in the correct direction towards \say{Likely} (42\% of the participants) and \say{Definitely Likely} (28\% of the participants). At the same time, most participants initially considered the true news items \say{Likely} and \say{Definitely Likely} to remain in these two levels. In addition, running a t-test on the distribution of the accuracy likelihood before and after the search for the true news task reveals a statistically significant difference (p-value $<$ 0.001). Therefore, results from Figure~\ref{accuracy_true} indicate that online search had a positive impact on the assessment of the accuracy of the true news headlines. 
 
In Figure~\ref{accuracy_false}, we note that before the online search, 96\% of the participants correctly assessed the presented news items as false. However, after browsing online, this rate decreases to 80\%. In fact, we observe that participants who initially classified the headline as \say{Definitely Unlikely} or \say{Unlikely}, moved in the wrong direction towards \say{Likely} and \say{Definitely Likely} after the search (19\%). However, running a t-test on the distribution of the accuracy likelihood before and after the search for the false news task reveals that the difference is not statistically significant (p-value $=$ 0.79). Therefore, we cannot conclude about the impact of the search on the accuracy assessment of participants. Nevertheless, observations from Figure~\ref{accuracy_false} suggest that online search might have misled some participants into believing false news items as true.

Overall, we conclude that the positive impact of online search for the assessment of true news items does not hold for false ones, suggesting that searching information online is more likely to lead students to validate true news than to refute false ones.

\vspace{-8pt}
\subsubsection{Change in general opinion}

\begin{figure*}[h]
\captionsetup[subfigure]{justification=centering} %
    \centering
      \begin{subfigure}{0.35\textwidth}
        \includegraphics[width=\textwidth]{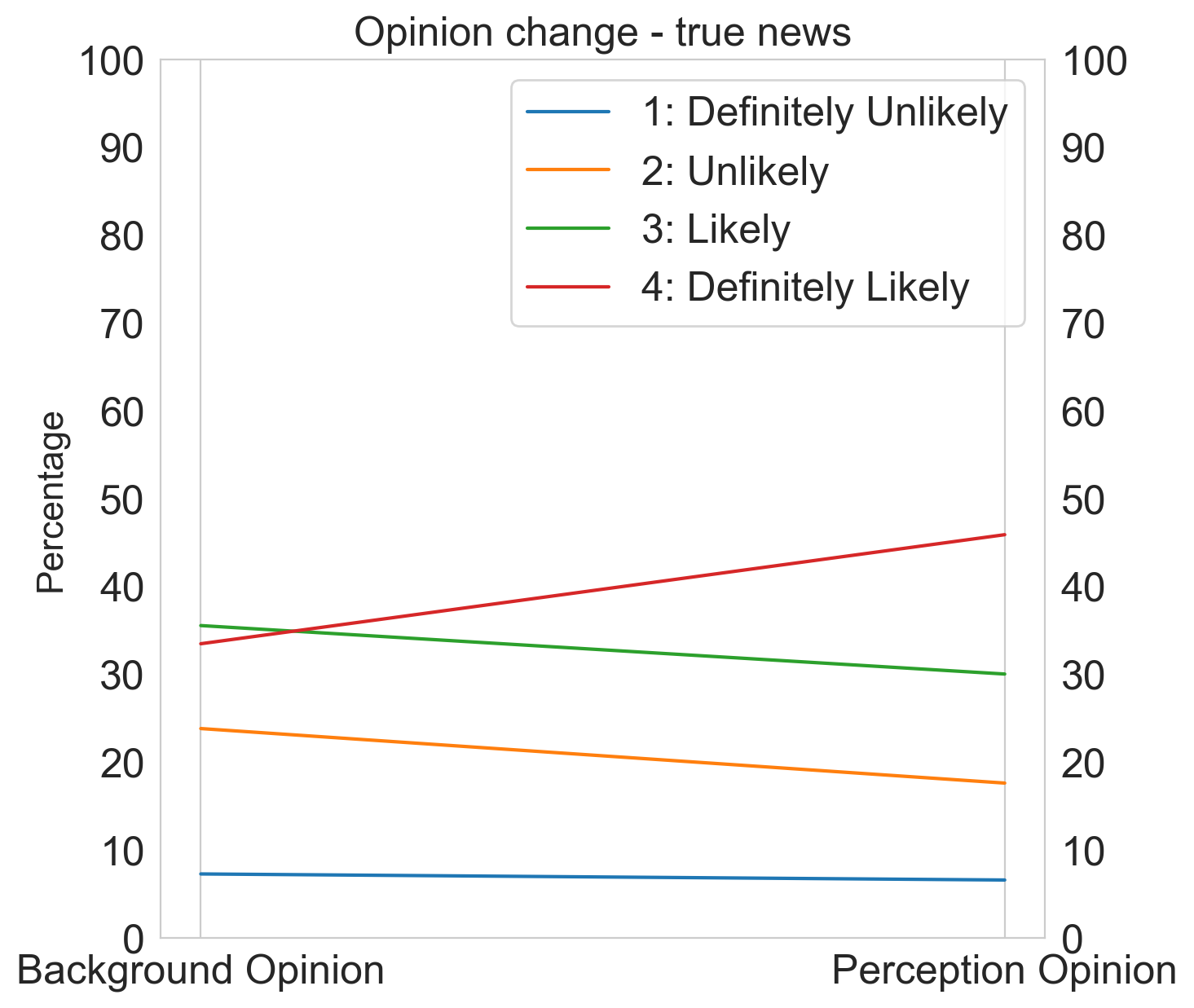}
        \vspace{-10pt}
          \caption{True News}
          \label{opinion_change_true}
      \end{subfigure}
      \begin{subfigure}{0.35\textwidth}
        \includegraphics[width=\textwidth]{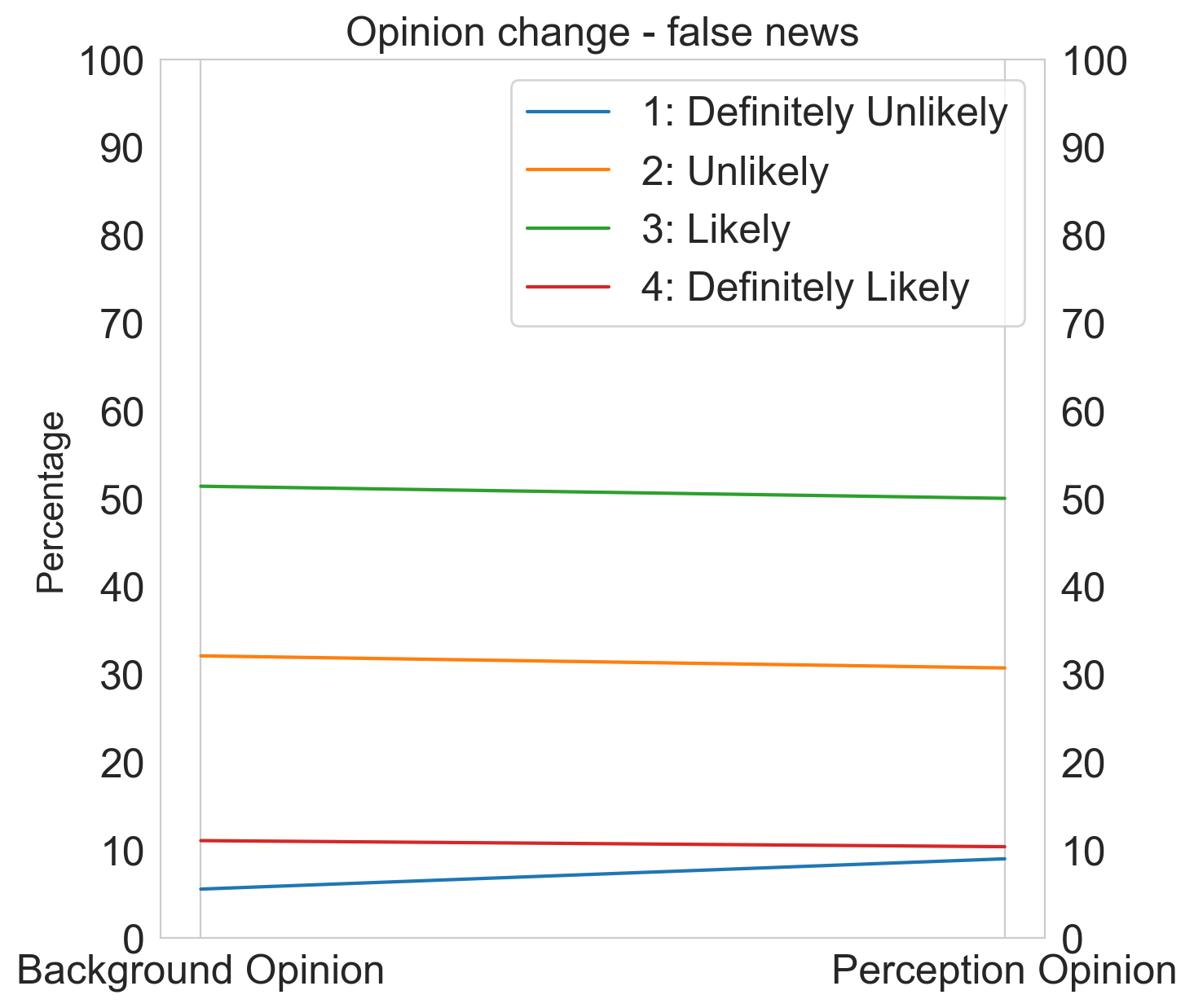}
\vspace{-10pt}
          \caption{False News}
          \label{opinion_change_false}
      \end{subfigure}
\vspace{-10pt}
      
\caption{Changes in participants’ general opinion about the topic between the \textit{Background} and \textit{Perception} for the true and false news. (a) True news: p-value~$<$~0.05. (b) False news: p-value~$=$0.33.}
\vspace{-15pt}

\label{opinion_change}
\end{figure*}

To develop our understanding of the impact of online search, we investigate the potential change in the \textit{general opinion} for both false and true news. To this aim, Figure~\ref{opinion_change} portrays participants’ change of general opinion between the Background and Perception phases (i.e., before and after Internet browsing). Specifically, Figure~\ref{opinion_change} presents the average general opinion score of topics related to true news (Figure~\ref{opinion_change_true}) and false news (Figure~\ref{opinion_change_false}). We observe that for the true news items, participants’ opinions shifted towards higher levels of accuracy (\say{Definitely Likely} and \say{Likely}) in the Perception phase (76\%) with respect to the Background phase (69\%) (p-value~$<$~0.05). When considering the topics connected to false news, participants’ opinions do not have a statistically significant difference across the two phases of the experiment (p-value~$=$~0.33).

Overall, for both true and false news topics, the amplitude of the changes in general opinion is small if compared to the changes in the accuracy judgment (\textit{cf.} Figure~\ref{accuracy_change}). This suggests a propensity among participants toward changing their assessment of a specific piece of information more readily than their broader position on a topic.

\vspace{-10pt}
\subsection{Accuracy success and participants characteristics (RQ2)}

In this section, to answer RQ2, we measure how good the participants are at distinguishing a true piece of information from a false one, and we inspect whether some characteristics of the participants correlate with their ability to classify news items accurately. In the following, we use the term \textit{Accuracy Success} when participants respond with an accuracy level equal or larger than 3 (\say{Definitely Likely} or \say{Likely}) to true news items and with an accuracy level equal or smaller than 2 (\say{Definitely Unlikely} or \say{Unlikely}) to false news items.
\begin{figure}[h]
    \centering
    \includegraphics[width=\columnwidth]{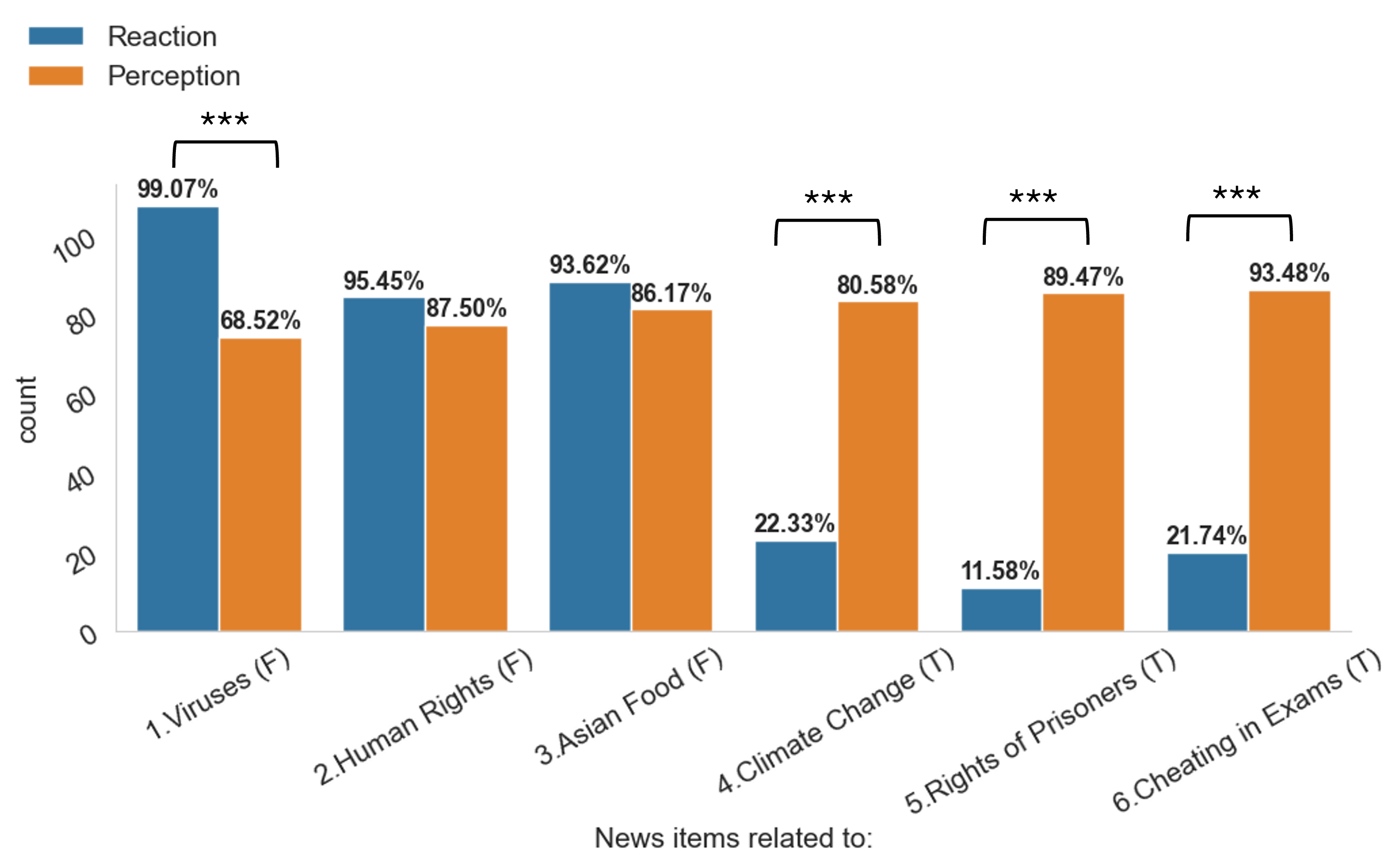}
    \vspace{-15pt}
    \caption{Accuracy Success distribution for the six different news items during the \textit{Reaction} and \textit{Perception} phases. Every pair of bars represents the topic of the news item along with its veracity within brackets (``T" stands for a true news item, whereas ``F" stands for a false news item). ***: p-value~$<$~0.001. There is no statistically significant difference between accuracy at \textit{Reaction} and \textit{Perception} phases for tasks 2 and 3.}
    \vspace{-15pt}
    \label{accuracy_success}
\end{figure}

In Figure~\ref{accuracy_success}, we present the Accuracy Success rate for both the Reaction and Perception phases (before and after browsing online). We observe a decrease in the success rate for false news items (related to tasks 1, 2, and 3). However, this decrease is only statistically significant for Task 1 (Covid-19),  where the drop is particularly noticeable. This might suggest that the abundance of information, and related infodemics, can negatively affect participants’ accuracy judgment on this topic. Contrarily, for true news items (related to tasks 4, 5, and 6), the success rate increase is remarkable (69\% on average, all increases are statistically significant). These observations further confirm that online search makes validating a piece of information easier than refuting it (see Section~\ref{sec4_1}). On average, the success rate across all the tasks in the Perception phase reaches almost 84\%, indicating that most participants are able to discern the veracity of a piece of information correctly. This observation is in line with the results obtained with adult participants in (\cite{pennycook2020fighting}, \textit{cf}. Study 2), which might suggest that the age of the participants does not impact their ability to recognize false information. 

To verify this hypothesis, we investigate possible relations between the characteristics of the participants and their ability to assess the veracity of news items successfully. To this aim, in Table~\ref{tab3}, we report the Pearson correlation scores between the Accuracy Success and the demographics of the participants collected in the profile survey. We find no correlation between any of the profile characteristics and the average Accuracy Success over the performed tasks\footnote{Please note that no correlation was found between the level of knowledge / interest of the participant in the topic and their accuracy judgment of the news item.}. 
In addition, we perform a regression analysis by using participants’ characteristics as independent variables and Accuracy Success as a dependent variable. However, this model can only describe 0.8\% of the variance in the dependent variable (R\textsuperscript{2} = 0.008). Indeed, demographic and rational thinking indicators are not reliable predictors (no significant coefficients) of participants' ability to correctly identify true and false news items. 

\begin{table}[h]
  \caption{Pearson correlation coefficients between the profile characteristics of the participants and the average Accuracy Success. None of the correlations is statistically significant.}
  \label{tab3}
  \begin{tabular}{cc}
    \toprule
                         & \textbf{Correlation with Accuracy Success} \\
                             \midrule
Age                  & 0.02                            \\
Gender               & -0.06                             \\
School grade         & -0.06                            \\
CRT score            & 0                              \\
RTL score            & 0.06                              \\
  \bottomrule
\end{tabular}
\end{table}
\vspace{-15pt}
\subsection{Information Sources (RQ3)}

\begin{figure}[h]
    \centering
    \includegraphics[width=\columnwidth]{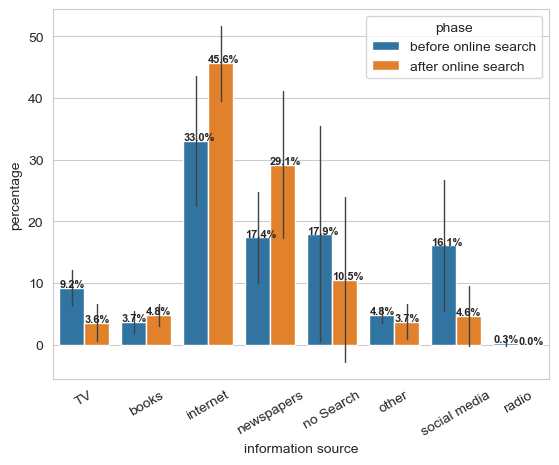}
    \vspace{-15pt}
    \caption{Distribution of information sources before and after online search.}
    \vspace{-10pt}
    \label{info_source}
\end{figure}

To answer RQ3, we examine the information sources the participants rely on and their impact on the Accuracy Success rate. In Figure~\ref{info_source}, we observe the distribution of participants’ favorite information sources. 
Before the online search, the main information sources are the \textit{Internet} (33\%) followed by \textit{newspapers} (17.4\%) and \textit{social media} (16.1\%). The percentage of participants who would not look for information on the topic is non-negligible (17.9\%).

After the online search, we notice a significant drop in the preferences of \textit{social media} as an information source (-11.5\%). While we observe an increase in the \textit{Internet} (+12.6\%) and \textit{newspapers} (+11.7\%). In addition, fewer participants would declare not searching for the topic as the \textit{no search} category drops by 7.4\%. From Figure~\ref{info_source}, we can also observe a high variance across the six tasks in the distribution of information sources, particularly for \textit{no search}, \textit{internet}, \textit{newspapers}, and \textit{social media}. This might be explained by the diverse nature of the presented topics and news, which in turn can elicit the preference of different resources to get information, as demonstrated by the different distributions across the six topics, displayed in the \textit{Appendix}. 

Overall, Figure~\ref{info_source} suggests that most participants choose to get information online and that many of them, after performing a search, chose not to get information from social media in the future.
\vspace{-5pt}
\subsubsection{Change of social media information source}

\begin{figure}[h]
    \centering
    \includegraphics[width=0.8\columnwidth]{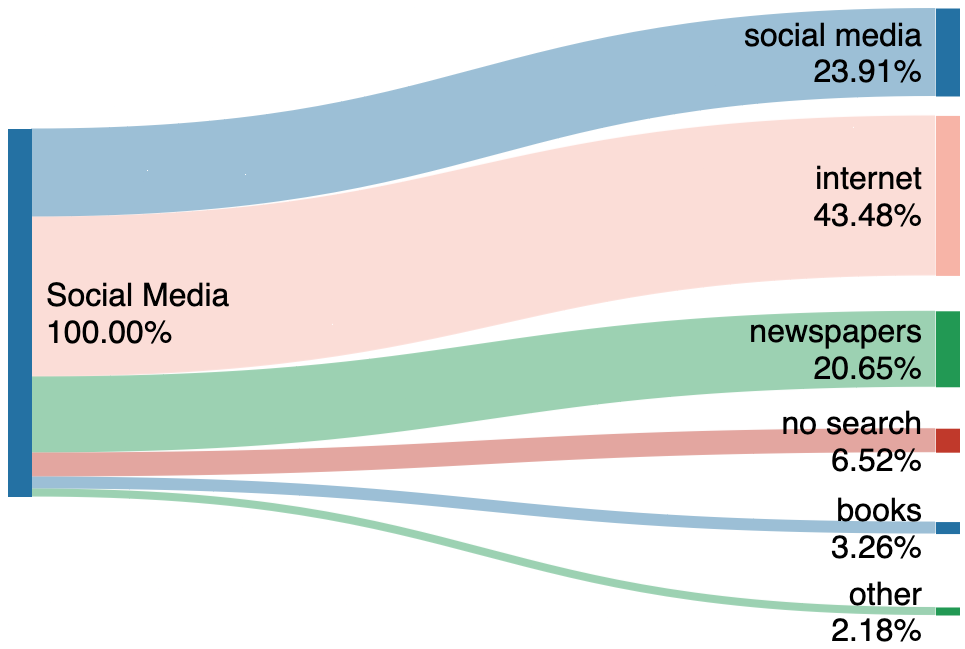}
    \vspace{-8pt}
    \caption{Distribution of information sources chosen during the Perception phase for the participants that answered social media to question 1.3 of the Background phase.}
    \vspace{-9pt}
    \label{social_media_change}
\end{figure}

In the following, we investigate the drop in the \textit{social media} information source in more depth. To this aim, we consider the 92 tasks in which users reported \textit{social media} as their favorite information source in the Background phase. We observe the change from social media to other information sources across the two phases in Figure~\ref{social_media_change}.
\vspace{-10pt}

We observe that the majority of changes (64\%) go from \textit{social media} to online resources such as \textit{Internet} and \textit{newspapers}. Almost 7\% of the participants that dropped \textit{social media} declared they will not search for information about these topics anymore, whereas 5\% moved to \textit{books} or \textit{other} sources. The remaining 24\% choose \textit{social media} for searches in the future. Note that none of the users that choose \textit{social media} in the Background phase selects \textit{TV} for future search. Figure~\ref{social_media_change} indicates that a fraction of participants considered social media unsuitable for searching for information, turning their preference to the \textit{Internet} for their information search in the future. 

\subsubsection{Online vs non-online information sources}
\label{sec4_1}

\begin{figure}[h]
    \centering
    \includegraphics[width=0.6\columnwidth]{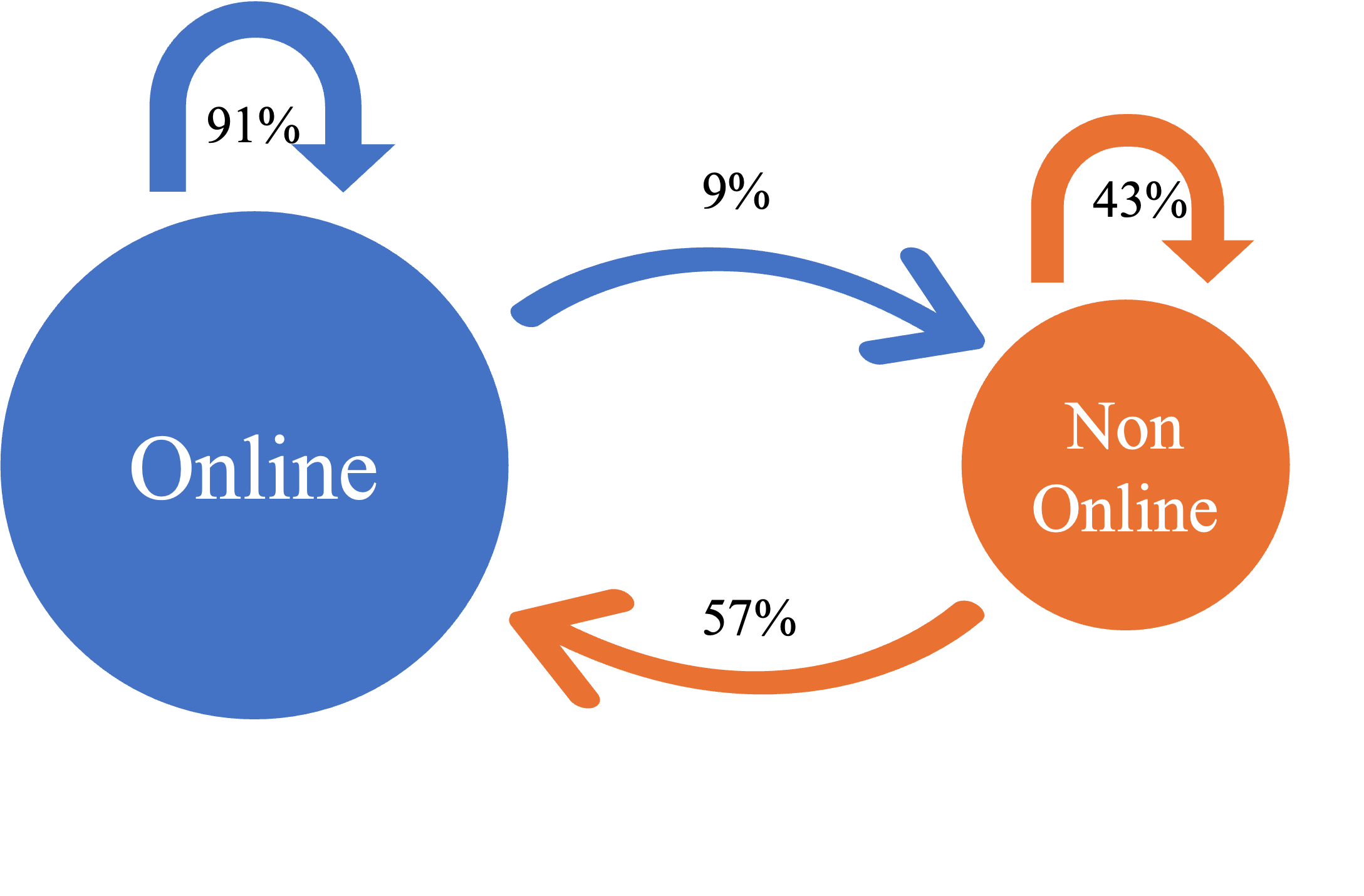}
    \vspace{-20pt}
    \caption{Change in online vs non-online information sources before and after browsing online.}
    \label{change_online}
    \vspace{-15pt}
\end{figure}

We explore more in depth the changes in information sources by grouping them into online sources (\textit{Internet}, \textit{newspapers}, and \textit{social media}) and non-online sources (\textit{radio}, \textit{TV}, \textit{books}, \textit{other}, \textit{I don't get/look for this information}). In Figure~\ref{change_online}, we present the changes in participants’ preferences between online and non-online information sources from the Background to the Perception phase by grouping the six tasks together. On the one hand, only 9\% of the participants who chose an online information source during the Background phase changed their preference in the Perception phase, choosing a non-online information source for future searches. On the other hand, almost half of the participants who initially declared that they were getting information from a non-online information source moved to online resource preferences. This indicates that, on average, participants tend to prefer online sources, and this preference is confirmed and shared by more people across all tasks after performing an online search. This might suggest that participants consider online searching a quick, more practical, and efficient way to find information about a specific topic.
\vspace{-8pt}

\subsubsection{Information source and Accuracy Success}
\begin{figure}[h]
    \centering
    \includegraphics[width=\columnwidth]{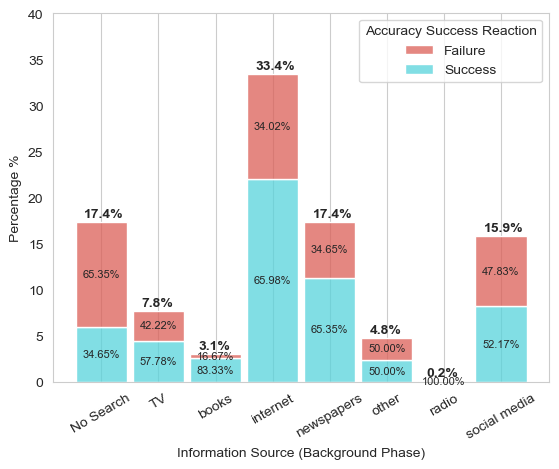}
    \vspace{-15pt}
    \caption{Information source at Background phase and Accuracy Success distribution at the Reaction phase. Percentages on top of the bars represent the distribution of the information sources among participants at the Reaction phase. Percentages within the bars represent the success rates for different information sources.}
    \vspace{-15pt}
    \label{info_source_success}
\end{figure}

To investigate whether there exists any relation between participants’ favorite information source and Accuracy Success, we portray, in  Figure~\ref{info_source_success}, the Accuracy Success rates at the Reaction phase for each information source from the Background phase. Information sources with the highest rate of Accuracy Success are \textit{books}, 83\% (15 participants), \textit{Internet}, 66\% (128 participants), and \textit{newspapers}, 65\% (66 participants)\footnote{We do not consider participants who get their information from the \textit{radio} (only one participant)}. Interestingly, \textit{social media} success rate is almost 13\% lower with respect to other online information sources (\textit{Internet} and \textit{newspapers}). The lowest success rate is related to participants who did not get information from any source (35\%).

\vspace{-8pt}
\section{Discussion \& Conclusions}
\label{conclusion}
In this paper, we examined how youth deal with the assessment of true and false pieces of information, inspecting several factors potentially affecting their accuracy judgment. By carrying out an experimental campaign with 183 participants, we gathered a novel dataset on youth online search behavior, interests, and knowledge and looked at changes in their opinions on six news items before and after an online search.

Our results suggest that online search helps verify a true claim rather than refuting a false one. Indeed, from our experiment, we noted a remarkable and statistically significant increase (of 69\%) in the accuracy success of true news items after performing an online search. At the same time, the accuracy success rate dropped for false news items (16\% drop). However, the former difference is not statistically significant. Therefore, we hypothesize that online search did not help participants in debunking false claims and could have misled some. These observations imply that simply searching for information online when encountered on social media, for example, could increase the accuracy perception of true news. 

Interestingly, we found no evidence of a relationship between participants’ demographics, rational characteristics, knowledge and interest levels, and accuracy judgment capabilities. We also found that participants are more likely to change their assessment of a specific news item than their opinion about a related broader topic. As for the information sources, the majority of participants declared to rely on online information sources before performing the online search, and this proportion increased even more after the online search. In addition, we observed a relevant shift from \textit{social media} to \textit{Internet} and \textit{newspapers} as favorite information sources across the different phases of the experiment. Overall, we found that, on average, young participants accurately discern a true piece of information from a false one 84\% of the time, and they are more likely to succeed if they took information from \textit{books}, \textit{Internet}, or \textit{newspapers}.

Through our research, we add valuable insights to the ongoing exploration of factors influencing the online news perceptions of young individuals. This work has the potential to build strategies to raise information literacy skills among vulnerable groups such as youth and to protect them from online threats such as misinformation, possibly curbing its diffusion.
\vspace{-6pt}
\paragraph{Limitations.} Our work has some limitations. First, after completing their first task, the participants might have guessed the purpose of the experiment. Hence, this might affect their performance in the following tasks. However, we controlled the order of task execution in our analysis, and no significant change related to the accuracy success was found. Second, some of the survey items could be misunderstood or misinterpreted. For example, the resource \textit{newspapers} can refer to both online and printed newspapers. Similarly, \textit{Internet browsing} can be interpreted in multiple ways, including online newspapers and social media pages. In addition, participants might not be sure of how to categorize multimedia platforms (e.g., Youtube), which can be considered as \textit{Internet browsing}, \textit{social media}, or even \textit{TV} in particular circumstances. Considering this, we suggest clearing up these ambiguities, including in the future the following choices: \textit{online newspapers}, paper-based \textit{newspapers}, and \textit{Internet browsing}: online forums, blogs, etc. Third, a control group with older participants could be used to identify potential differences with respect to youth. By contrasting our results with those in (\cite{pennycook2020fighting}, \textit{cf.} Study 2), we commence a first comparison that needs to be expanded in future work.
\vspace{-5pt}
\paragraph{Future Efforts.} Regarding our next endeavors, we plan to analyze the browsing logs of the participants, which we collected during the online search phase of the experiment, and whose analysis was not within the scope of this paper. We aim to extract, if existent, particular search patterns and identify personal search styles among participants. Through this analysis, we also plan to investigate the use of LLMs in search and their impact on accuracy assessment. Additionally, we intend to leverage Natural Language Processing (NLP) techniques to perform an analysis of the search queries performed by the participants, which might enable the prediction of the task’s success or failure. Moreover, we plan to include more participants in the experiment, which can help build stronger conclusions on the effect of search on the assessment of false news. In fact, with a small sample, type II errors can easily increase. By increasing the sample size, we will be more likely to observe statistically significant results.
\vspace{-5pt}
\paragraph{Ethical Considerations.} 
The participants in the experiment were duly informed about the collection of their demographic information from the profile survey and their navigation actions and answers from the task surveys. The analysis presented in this paper was conducted on data anonymized by design, which was stored on secure servers only accessible by the researchers involved in this paper. 
\section*{Acknowledgments}
The authors thank Dr. Zachary J. Roman for the helpful discussion. This work is partially funded by the Swiss National Science Foundation grant 100019\_188967. 

\bibliography{aaai22}

\appendix

\section*{Appendix}
\begin{figure}[h]
  \centering
  \label{fig_headline}
  \includegraphics[width=\linewidth]{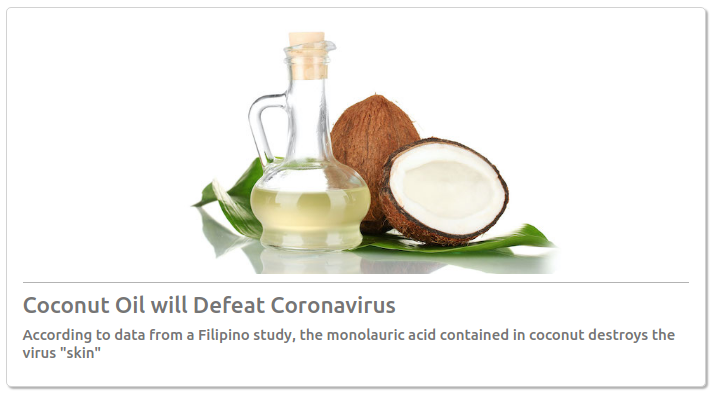}
  \caption{Example of the headline presented to the participants during Task 1.}
\end{figure}

\begin{figure}[h]
    \centering
    \includegraphics[width=\columnwidth]{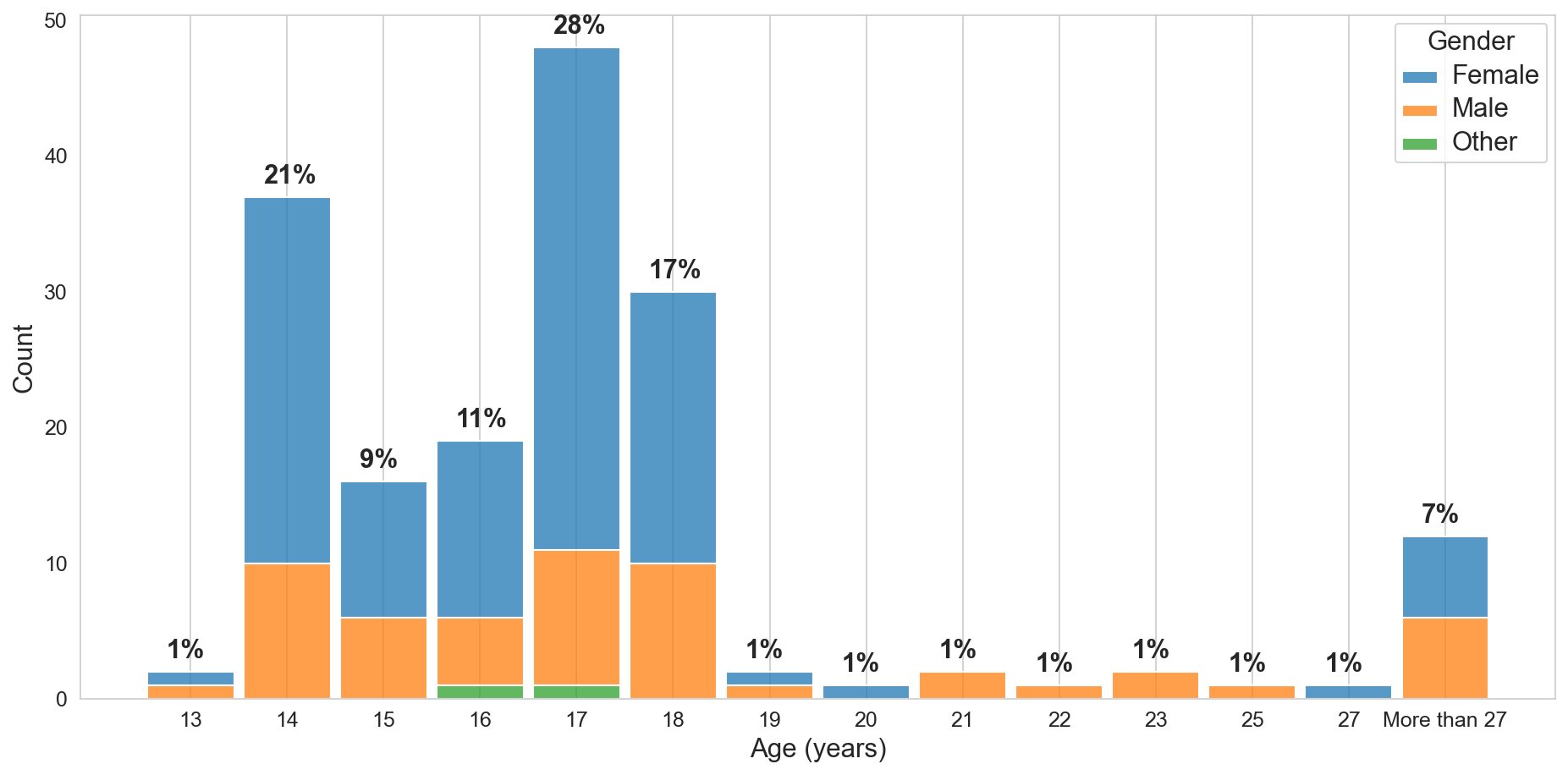}
    \vspace{-10pt}
    \caption{Age distribution of participants.}
    \label{fig_Age_dist}
\end{figure}

\begin{figure}[h]
    \centering
    \includegraphics[width=\columnwidth]{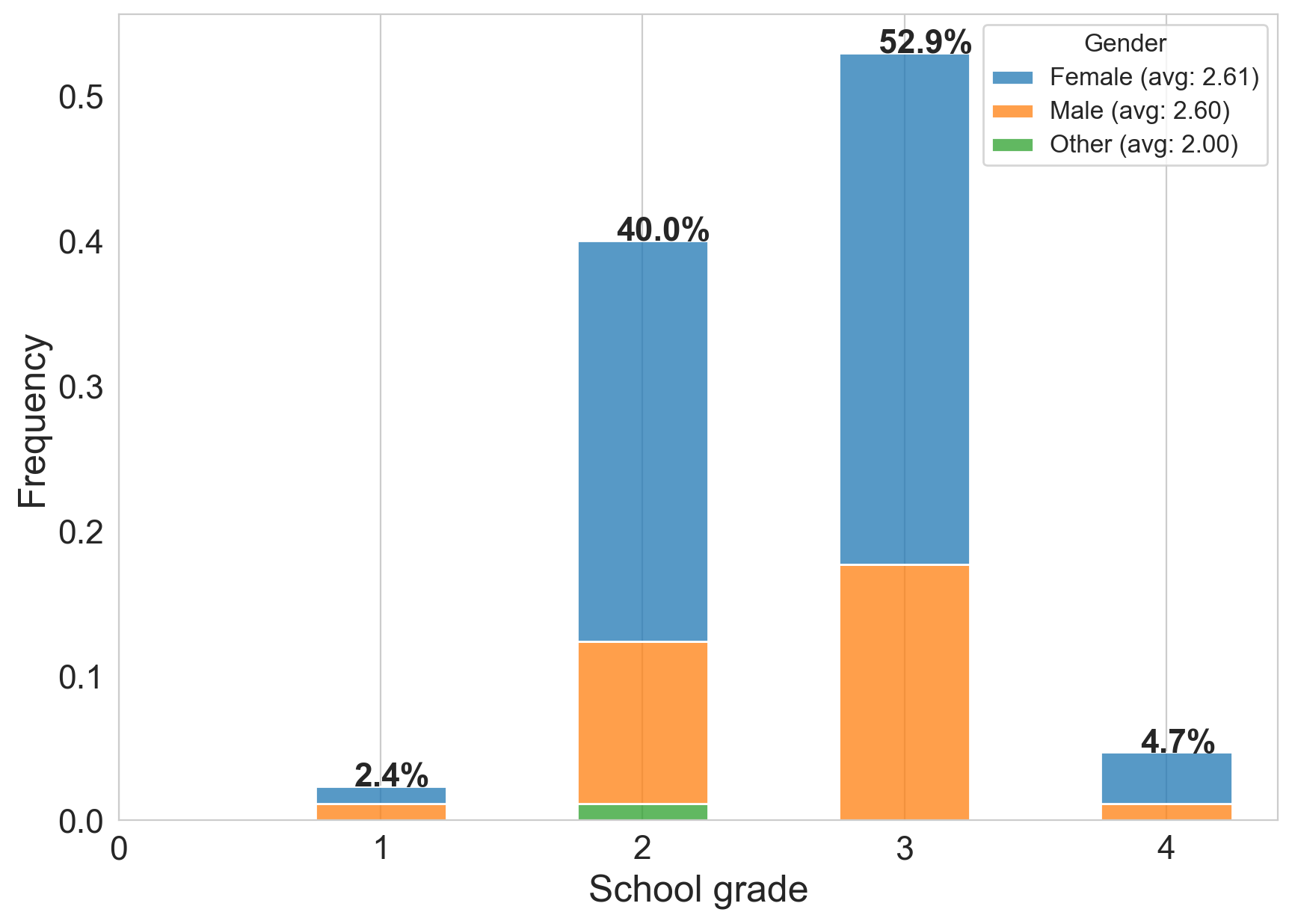}
    \vspace{-10pt}
    \caption{School grades distribution of participants by gender.}
    \label{fig_grades}
\end{figure}

\begin{figure}[!h]
    \centering
    \includegraphics[width=\columnwidth]{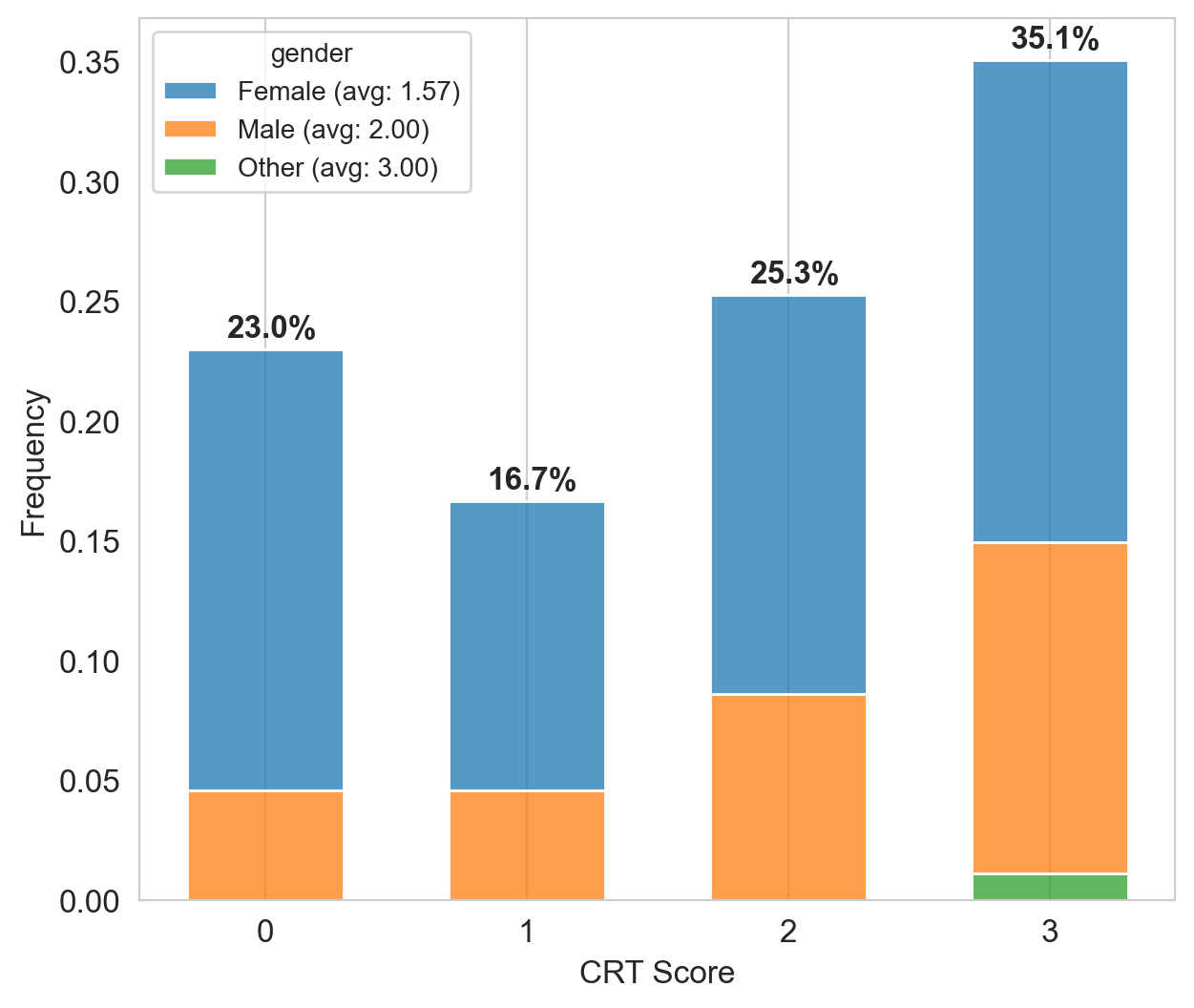}
    \vspace{-10pt}
    \caption{Participant’s Cognitive Reflection Test (CRT) score distribution.}
    \label{fig_CRT}
\end{figure}

\begin{figure*}[t]
    \centering
    \includegraphics[width=\textwidth]{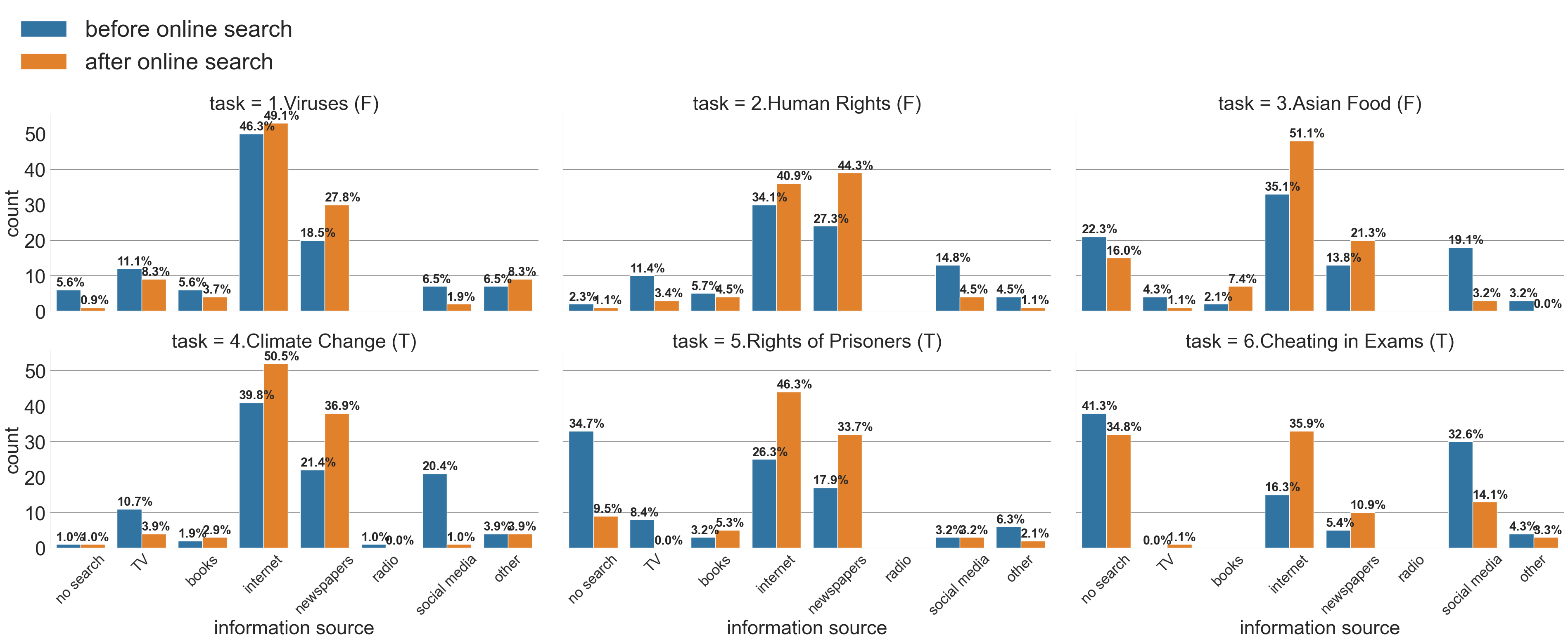}
    \caption{Distribution of information sources for all tasks before and after performing online search. "T" stands for a true news item, whereas "F" stands for a false news item.}
    \label{fig_Info_source_all}
\end{figure*}

We present in the following some additional information from the experiment. In Table~\ref{tab_users_search} the Distribution of users that did the search for the six tasks. An example of headline presented to the participants during the \textit{Reaction} phase of the \textit{task survey} is in Figure~\ref{fig_headline}. The age distribution of the participants is presented in Figure~\ref{fig_Age_dist}. The school grade distribution is shown in Figure~\ref{fig_grades} (Scale: 1 = insufficient, 2 = Sufficient, 3 = good, 4 = very good). The Cognitive Reflection Test score distribution is reported in Figure~\ref{fig_CRT}. In addition, Table~\ref{tab4} contains details on the six tasks of the experiment, and Figure~\ref{fig_Info_source_all} the information source distribution for every task.

\begin{table}[h]
\caption{Distribution of users that did the search for the six tasks.}
\label{tab_users_search}
\begin{tabular}{p{0.06\columnwidth}p{0.34\columnwidth}p{0.34\columnwidth}p{0.1\columnwidth}}
\toprule
\textbf{Task} & \textbf{Number of users that did the task survey} & \textbf{Number of users that did the search} & \textbf{search rate}  \\ \midrule
1    & 243                                     & 109                                & 44.86\%     \\
2    & 242                                     & 88                                 & 36.36\%     \\
3    & 240                                     & 87                                 & 36.25\%     \\
4    & 240                                     & 103                                & 42.92\%     \\
5    & 246                                     & 94                                 & 38.21\%     \\
6    & 244                                     & 89                                 & 36.48\%    \\ \bottomrule
\end{tabular}
\end{table}

\vspace{-10pt}


\begin{table*}[t]
\caption{Headlines and leads of the six tasks presented in the survey.}
\label{tab4}
\begin{tabular}{@{}p{0.05\textwidth}p{0.05\textwidth}p{0.15\textwidth}p{0.15\textwidth}p{0.5\textwidth}@{}} 
\toprule
\textbf{ID} & \textbf{Type}  & \textbf{Topic}                  & \textbf{Headline}                                                             & \textbf{Lead}                                                                                                                     \\ \midrule
1  & False & Virus                  & Coconut oil will beat Coronavirus                                        & According to data from a Filipino study, the monolauric acid contained in the coconut destroys the cover of the virus        \\
2  & False & Human Rights Violation & Covid concentration camps for \say{disobedient} Canadian citizens            & A parliamentary hearing reveals the state of Ontario's secret plan to deal with the crisis                                   \\
3  & False & Asian Food             & The new McDonald's dog burger conquers Korea                             & McDog is the name of the new menu designed to satisfy the tastes of Korean customers                                         \\
4  & True  & Climate Change          & Malaria spreads in Europe due to climate change                          & Due to rising temperatures, the mosquitoes that spread the virus are on the doorstep of Europe.                              \\
5  & True  & Prisoners rights       & Guards torture prisoners with Baby Shark                                 & Five inmates were tortured with the well-known children's song in Oklahoma, and three former prison guards are now on trial. \\
6  & True  & Cheating in exams       & Son disguises himself as his mother to get her to get a driver's license & Wearing a long skirt and padded bra, the 43-year-old Brazilian showed up for his driver's license exam. But it did not work. \\ \bottomrule
\end{tabular}
\end{table*}

\end{document}